\documentclass[final,3p,times,twocolumn]{elsarticle}

\usepackage{graphics}
\usepackage{graphicx}
\usepackage{epsfig}

\usepackage{amssymb}
\usepackage{amsthm}

\usepackage{amsmath}

\usepackage{color}

\newcommand\btau{\mbox{\boldmath$\tau$}}
\def\slash#1{\ooalign{\hfil/\hfil\crcr$#1$}}

\journal{Physics Letter B}

\begin{document}

\begin{frontmatter}



\title{Variational approach to the inhomogeneous chiral phase in quark matter}


\author{S.Karasawa and T.Tatsumi}

\address{Department of Physics, Kyoto University, Kyoto 606-8502, Japan}

\begin{abstract}
The inhomogeneous chiral phase is discussed in QCD at finite temperature and/or density. 
We study the phase diagram on the density-temperature plane by taking into account the effect of the current mass 
by a variational method.
It is demonstrated that our framework well describes the inhomogeneous phase over the whole phase region.
\end{abstract}


\end{frontmatter}



Recently the inhomogeneous chiral phases have been actively studied in the QCD phase diagram; where the chiral condensates (order parameter) have a spatial dependence. 
Inhomogeneous order parameter has been studied as well in condensed matter physics; e.g., spin density wave \cite{SDW1,SDW2}, charge density wave \cite{CDW}, Fulde-Ferrell-Larkin-Ovchinikov superconductor \cite{FFLO1,FFLO2} and so on. 
We here consider the similar subject about chiral transition within QCD. 
It has been shown in \cite{BD,BDT} that there are exact solutions of inhomogeneous chiral condensates
 in 1+1 dimensions and inhomogeneous phases actually develops up to a critical temperature. 
It has been shown that analytic solutions are no longer real functions but complex functions such as {\bf $D(z) = \langle \bar{\psi} \psi \rangle +i \langle \bar{\psi} i \gamma_{5} \tau_{3} \psi \rangle =\Delta(x) e^{i \theta(x)}$}, 
and represented by the Jacobi elliptic functions in the two dimensional NJL model in the large $N_c$ limit. 
In \cite{Nickel}, it has been shown that one dimensional inhomogeneous structure can be embedded in 1+3 dimensions. 
He showed that the energy spectrum of quarks in the 1+1 dimensional systems can be generalized to 1+3 dimensions by operating the Lorentz boost in the perpendicular direction to the spatial modulation.
He then applied such a method for the real kink crystal (RKC), which has a spatial dependence in the amplitude of $D(z)$; $D(z) = \Delta(z)$, and has shown that the inhomogeneous chiral phases may appear in the low temperature and/or moderate density region before chiral transition in the QCD phase diagram in 1+3 dimensions. 
Similar procedure can be applied for another type of condensations. 
Actually, prior to \cite{Nickel}, \cite{Nakano-Tatsumi} have suggested that dual chiral density wave (DCDW), which has a spatial dependence in the phase of  
$D(z)$; $D(z) = \Delta e^{i\theta(z)}$, also appears in the QCD phase diagram within the 1+3 dimensional NJL model. 

Nevertheless, many studies have considered only idealized situations such as no magnetic field, no isospin asymmetry, and no current quark mass (chiral limit).
External magnetic field and isospin asymmetry are especially important in relation to realistic implications on the heavy ion collision or compact-star phenomena. 
Since the magnitude of the current quark mass is not more than a few MeV, one may consider that it can be neglected. However, it actually makes inhomogeneous phases more enriched. 
The current quark mass is indeed small compared with the constituent quark mass
 but the mass term breaks chiral symmetry explicitly, so the Nambu-Goldstone boson becomes massive. 
This phenomena is important in the low temperature and/or density region, where
 the physical pion mass of 138 MeV is never small. 
Since the inhomogeneous phases appear in the moderate
 density region (around $\mu =$ 300-400 MeV), the pion mass can not be neglected there.
Moreover, in the phase boundary of chiral transition (critical point)
, where the constituent quark mass become very small and comparable with the pion mass, we can expect the importance of 
the current quark mass.
Actually, it has been demonstrated that the properties of the chiral transition have been much modified not only quantitatively but also qualitatively by the current mass \cite{Buballa}. 

Once the current quark mass have been switched on, function form of inhomogeneous order parameters
 will be changed from the one in the chiral limit.
We here study the change of order parameters and figure out its consequences in the QCD phase diagram. 
Exact solutions to the RKC-type inhomogeneous structure in the massive case has been already studied in \cite{RKC-mass}, while there have been little studies about the DCDW-type structure. 
This is because the exact solutions have not been obtained for the DCDW-type structure in the presence of the current quark mass, 
while this configuration bears rich physical contents and might have phenomenological implications as we will see latter.




One way to discuss such effect is a perturbative approach from the chiral limit, without changing quark wave function and the form of inhomogeneous chiral condensates. 
This approach has been discussed in \cite{Maedan}, but may not be sufficient to study the DCDW-type structure. 
As we shall see, the function form of the inhomogeneous order parameters has been largely changed, especially near the critical point.


Note that the current quark mass effect to the DCDW-type inhomogeneous structure changes it qualitatively while the RKC-type changes it quantitatively.
DCDW utilizes the chiral surface after spontaneous symmetry breaking, which is formed by the minima of the effective potential; all the points on the surface, designated by $\left( \langle{\bar \psi}\psi \rangle, \langle{\bar \psi} i \gamma_5 \boldmath{\tau} \psi \rangle \right)$, are then equivalent to each other in the chiral limit.
Usually only the point $\left( \langle {\bar \psi} \psi \rangle , {\bf 0} \right)$ is chosen as a fiducial point.
Restricting the isospin degrees of freedom to $\tau_3$, we have a chiral circle $U(1)$ specified by the chiral angle $\theta$.
DCDW uniformly winds along the chiral circle with the spatial dependence of $\theta$, $\theta(z)=qz$; the energy increases in the kinetic energy is then compensated by the attractive interaction between DCDW and quarks.
Once the mass term is taken into account, the chiral circle deforms and points on the chiral circle become no more equivalent to each other for the point $\theta=0$ to be the absolute minimum.
Accordingly $\theta(z)$ should be deformed to take an optimal form.
This is the leading order effect and qualitatively modifies the form of $\theta(z)$.
Note that the amplitude of DCDW should not be so affected by the inclusion of the finite quark mass, since the modification of the radius of the chiral circle should be a minor effect.
Since the spatial dependence is restricted to the radius direction for RKC, $ \mathbb{Z}_2$ symmetry is violated in this case. 
Then, RKC should be changed only quantitatively by the finite quark mass effect.
This is the reason for the importance of considering effect of the finite quark mass in the DCDW-type inhomogeneous structures.


To look for the unknown DCDW-type inhomogeneous structure, we propose here a variational approach. 
We start from two flavor and three color NJL model in 1+3 dimensions at finite temperature and chemical potential. 
Here we assume no external magnetic field and no isospin asymmetry. Lagrangian is given as
\begin{align}
\mathcal{L}_{\mathrm{NJL}} = \bar{\psi}( i \slash \partial -m_c )\psi +G[(\bar{\psi}\psi)^{2}+(\bar{\psi} i \gamma_{5} \btau \psi)^{2}] , \label{0}
\end{align}
where $m_c$ is the current quark mass.
The form of the chiral condensates is assumed such that 
\begin{align}
& \langle \bar{\psi} \psi \rangle = \Delta \cos \theta(z) ,  \label{1} \\
& \langle \bar{\psi} i \gamma_{5} \tau_{3} \psi \rangle = \Delta \sin \theta(z) , \label{2} 
\end{align}
as a generalization from DCDW in the chiral limit: 
$\Delta$ is a uniform amplitude of condensates, and the chiral angle $\theta(z)$ is now a dynamical variable, and reduced to $qz$ in the chiral limit. 
Using the mean-field approximation and inserting Eq.(\ref{1}) and Eq.(\ref{2}) in the NJL Lagrangian Eq.(\ref{0}), the Lagrangian reads
\begin{align}
\mathcal{L}_{\mathrm{MF}} = \bar{\psi}[ i \slash \partial - M - U(\theta(z)) ]\psi - G \Delta^2, \label{3}
\end{align}
where $M = m_c - 2G \Delta$ is the dynamical quark mass 
and $U(\theta(z)) = -2G \Delta \exp (i \gamma_5 \tau_3 \theta(z)) - M + m_c$.

The mean-field thermodynamic potential then renders
\begin{equation}
\Omega_{\rm MF} = - T {\rm{Tr}} \log [ i \slash \partial - M + \mu \gamma_0 - U(\theta(z)) ] + G \Delta^2 V,
\end{equation}
where the operator trace ($\rm{Tr}$) means the sum over flavor, color, Matsubara frequency and integration over the momentum space. 
$V$ is the volume of the system,
Our main purpose here is to determine the inhomogeneous condensates and quark wave function self-consistently.
Introducing the Dirac operator
\begin{align}
H_{\rm{D}} = -i \gamma_0 \mbox{ \boldmath $\gamma \cdot \bigtriangledown$ } + \gamma_0 M  + \gamma_0 U(\theta(z)),
\end{align}
we explicitly write the self-consistent equations
\begin{align}
&H_{\rm{D}} \psi_{\alpha} = E_{\alpha} (\theta) \psi_{\alpha},  \label{a1} \\
& \langle \bar{\psi} \psi \rangle +i\langle \bar{\psi} i \gamma_{5} \tau_{3} \psi \rangle= \Delta e^{i \theta(z)}. \label{a2}
\end{align}
If we can solve Eqs.(\ref{a1}), (\ref{a2}) simultaneously, we would find the general inhomogeneous structure and their thermodynamic potential, 
but it might be too difficult to solve them analytically. Instead, we propose an alternative method here.
First we intuitively guess the function form of the chiral angle $\theta(z)$ by referring to  
the 1+1 dimensional case, where it has been shown that $\theta$ can be well approximated by the solution of the sine-Gordon equation in the presence of the mass term \cite{Schon-Thies}. 
We can then expect that the one-dimensional structure of the similar kind should be also realized in our case by embedding it in 1+3 dimensions. 
Actually we can see that such equation emerges by way of the derivative expansion of $\Omega_{\rm MF}$ 
and the solutions consist of the one-parameter family. Then we solve the Dirac equation for quarks using these solutions. 
We shall see this procedure works well in the whole region of the inhomogeneous chiral phase. 

Rewriting $\Omega_{\rm MF}$ in the form, $\Omega_{\rm MF}=-T{\rm Tr}{\rm log}\left[S_0^{-1}-U\right]$, 
with $S_0$ being the free propagator of quarks at finite $T,\mu$,
\begin{align}
&S_0 (x,y) = \frac{1}{i \slash \partial + \mu \gamma_0 - M} \delta(x-y).  \label{7} 
\end{align}
We invoke the derivative expansion; expanding $\Omega_{\rm MF}$ with respect to $\partial$ in a Taylor series;
\begin{align}
{\rm Tr } \log \left[ S_0^{-1} - U(\theta (z)) \right] &= {\rm Tr} \log \left[ S_0^{-1} ( 1 - S_0 U(\theta (z))) \right] \notag \\
&\sim \sum_{n=1}^{\infty} \frac{1}{n} \left( \frac{1}{i \slash \partial + \mu \gamma_0 -M} U(\theta(z)) \right)^n . \label{derivative}
\end{align}
Here we have neglected ${\rm Tr} \log S_0^{-1}$ because 
it has no influence on the equation of motion (EOM) of $\theta(z)$.
We can regard Eq. (\ref{derivative}) as infinite sum of Feynman diagrams and rewrite with respect to derivative, we find  
\begin{align}
\Omega_{\mathrm{MF}}^{\rm der} = \frac{1}{2} f_{\pi}^{*2} \left( \partial_{z} \theta(z) \right)^2 - m_c \Delta \cos \theta(z), \label{lde}
\end{align}
up to $\mathcal{O}(\partial^2)$ \cite{DE1,DE2}. 

If $|d\theta/dz|$ becomes large, higher derivative terms may become important.  
However, the mass term becomes irrelevant there and the solution should always be the linear function as in the original DCDW, 
which can be reproduced  
only by the leading term.
Here we drop the irrelevant constants for deriving the EOM of $\theta(z)$, and $f_{\pi}^*$ is the pion decay constant in medium defined by;
\begin{align}
&f_{\pi}^{*2} = \frac{N_f N_c M^2}{16 \pi^3} \Bigg[ 2 \pi {\rm{Ei}}(-M^2) - \sum_{s=\pm} \int d^3p \frac{1}{E^{3/2}} \frac{1}{1+e^{\beta (E+s\mu)}}  \Bigg],
\end{align}
where ${\rm{Ei}}(x)$ is the exponential integral, $E=\sqrt{p^2+M^2}$ and $\beta$ is the inverse temperature; $\beta = 1/T$.

Then we derive the EOM of $\theta(z)$ from the variational principle, $\delta \Omega_{\rm{MF}}^{\rm der} / \delta \theta = 0$, 
\begin{align}
\frac{d^2 \theta(z)}{dz^2} - m_{\pi}^{*2} \sin \theta(z) = 0 , \label{9}
\end{align}
where we used the Gell-Mann-Oakes-Renner relation, $f_\pi^{*2}m_\pi^{*2}=-m_c\Delta$. Eq.(\ref{9}) is the sine-Gordon (SG) equation. 
Obviously the SG equation is reduced to the massless Klein-Gordon equation in the chiral limit, 
and the chiral angle is then given by the linear function, which gives the trigonometric function 
for the condensates through Eqs.(\ref{1}) and (\ref{2}). This configuration is the original DCDW. 
If $|d\theta/dz|\gg m_\pi^{*}$, the second term can be safely neglected. Thus the original DCDW is an good approximation there.
On the other hand, it becomes important as $|d\theta/dz|\simeq m_\pi^{*}$; we can expect a large deformation of the original DCDW there.
Note that we have only used the free quark
propagator in this derivation. We shall take into account 
the back-reaction of $\theta$ or condensates to the quark wave-function in the next step.

The solution of the SG equation is given by the Jacobian amplitude function ${\rm am} (x,k)$:
\begin{align}
\theta (z) = \pi + 2 {\rm am} \left( \frac{m_{\pi}^* z}{k} , k \right) , \label{10}
\end{align}
where $0 \leq k \leq 1$ is the modulus parameter. Note that $k$ measures the deformation from DCDW; 
in the limit, $k \rightarrow 0$, ${\rm am}(x,k) \rightarrow x$ or $\theta(z)$ is reduced to a linear function:
\begin{align}
\theta(z) \rightarrow \pi + \frac{2m_{\pi}^*}{k}z ~~~{\rm when}~ k \rightarrow 0 . \label{11}
\end{align}
This means that the inhomogeneous condensates are reduced to the original DCDW when $k \rightarrow 0$.
In the opposite limit,  $k \rightarrow 1$, ${\rm am}(x,k) \rightarrow \arcsin \tanh(x)$ so that $\theta(z)$ goes to sine-Gordon kink:
\begin{align}
\theta(z) \rightarrow \pi + 2 \arcsin \tanh \left( \frac{m_{\pi}^*z}{k} \right) ~~~{\rm when}~ k \rightarrow 1 . \label{12}
\end{align}
Taking together, we can see that as a value of $k$ becomes larger, the spatial modulation of the chiral angle is more remarkable.

Once the function form of $\theta$ is obtained, we try to solve the Dirac equation (\ref{a1}) in the background of $\theta$. 
However, it is a hard task yet. 
To circumvent the difficulty we take the following procedure.
Using the property, $d\theta(z)/dz=(2m_\pi^*/k){\rm dn}(m_\pi^*z/k,k)$, we can see that it fluctuates around some mean value. Thus 
we may expect that the DCDW-type wave function should be suitable for the quark field by 
 replacing the derivative term by its spatial average. We write $\psi={\cal N}{\rm exp}(i\gamma_5\tau_3 qz/2){\exp(i{\bf p}\cdot{\bf r})}$, 
for momentum $\bf p$ with a variational parameter $q$, which is then related to $k$ by the relation, 
\begin{align}
&q \equiv \langle \partial_z \theta(z) \rangle = \frac{\pi m_{\pi}^*}{k{\bf K}(k)}, \label{qfomula}
\end{align}
with the complete elliptic integral of the first kind,  ${\bf K}(k)$ . Substituting the DCDW-type wave function in Eq.~(\ref{a2}), 
we can easily see the self-consistency is satisfied if the spatial average of $\theta$ is taken in R.H.S.
We will examine the validity of our variational ansatz in the numerical results to see it works over the whole region of the inhomogeneous chiral phase.

Finally the thermodynamic potential is given as 
\begin{align}
&\Omega_{\rm{MF}} \simeq \Omega_{\rm{DCDW}} + \Omega_{\rm{SB}} + G \Delta^2 V, \label{6} \\
&\Omega_{\rm{DCDW}} = -T {\rm{Tr}} \log [S_{\rm{ref}}^{-1} + \mu \gamma_0], \label{7} \\
&\Omega_{\rm{SB}} = -T {\rm{Tr}} [(S_{\rm{ref}}^{-1} + \mu \gamma_0 )^{-1} F(m_c ; \theta(z))]. \label{8},
\end{align}
where $S_{\rm ref}$ is the DCDW-type propagator, $S_{\rm{ref}}^{-1} (\theta(z)) = i \slash \partial - M + q/2 \tau_3 \gamma_5 \gamma_3 $.
and the remaining $O(m_c)$ term, $F(m_c ; \theta(z)) = -m_c (\cos (\theta(z)) -1 + i \gamma_5 \tau_3 \sin (\theta(z)))$. 
Here $\Omega_{\rm{DCDW}}$ is the formula of the DCDW thermodynamic potential with the wave number $q$ and $\Omega_{\rm{SB}}$ represents the effect of explicit symmetry breaking of $O(m_c)$. 
Note that in the chiral limit $\theta(z) \rightarrow qz$ and $\Omega_{\rm{SB}} \rightarrow 0$ so that Eq.(\ref{6}) recovers the original DCDW \cite{Nakano-Tatsumi}. 
Inserting the form of $\theta(z)$ Eq.(\ref{10}) in the thermodynamic potential Eq.(\ref{6}), we explicitly write down the thermodynamic potential:

\begin{align}
&\Omega_{\rm MF} = \Omega_{{\rm DCDW}} + \Omega_{\rm SB} + \frac{(M - m_c)^2}{4G} , \label{13} \\
&\Omega_{\rm DCDW} = \Omega_{\rm vac} + \Omega_{\rm thermal},  \notag \\
&\Omega_{\rm vac} = \frac{N_f N_c}{8{\pi}^{5/2}} \sum_{s = \pm } \int_1^{\infty} \! \frac{dx}{x^{5/2}} \int_{0}^{\infty} dy \left[ e^{-(\sqrt{y^2+M^2}+sq/2)^2x} \right],  \notag \\
&\Omega_{\rm thermal} = -\frac{N_f N_c T}{2\pi^2} \sum_{s,l = \pm} \int_{0}^{\infty} \! dy \int_{0}^{\infty} \! dz ~y \notag \\
&~~~~~~~~~~~~~~~~~~~~~~~ \Biggl[ \log \left\{ e^{- \beta ((y^2 + (\sqrt{z^2+M^2}+sq/2)^2)^{1/2}+l\mu )} +1 \right\} \Biggr], \notag \\
&\Omega_{{\rm SB}} = N_f N_c f_{\pi}^{*2} m_{\pi}^{*2} \frac{2}{k^2 {\bf K}(k)} \left[ {\bf E}(k) - \left( 1-k^2 \right) {\bf K}(k) \right], \notag
\end{align}
where ${\bf E}(k)$ is the complete elliptic integral of the second kind.
$\Omega_{\rm DCDW}$ is the thermodynamic potential of the original DCDW; 
$\Omega_{\rm vac}$ is the contribution from the vacuum and $\Omega_{\rm thermal}$ is the contribution from finite temperature and chemical potential.

\begin{figure}[h]
\includegraphics[scale=0.3]{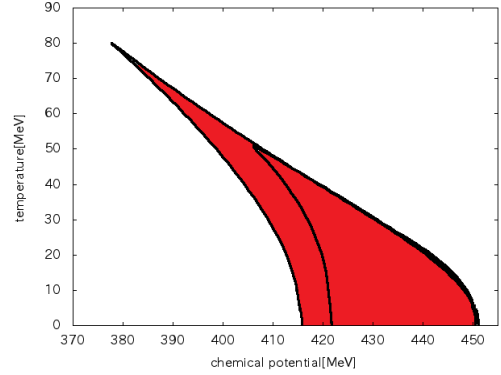}
\caption{QCD phase diagram at finite temperature and/or density in $m_c=0,5$MeV. Red region is the chiral inhomogeneous phase. As the current quark mass increases, the area of inhomogeneous phase monotonously decreases.}
\label{ddcdw1}
\end{figure}

\begin{figure}[h]
\includegraphics[scale=0.55]{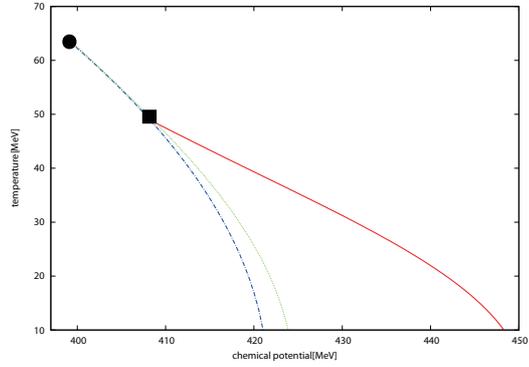}
\caption{QCD phase diagram in $m_c=5$MeV. Blue (dash) line denotes the phase boundary of the left hand side of the inhomogeneous phase. Red (solid) line also denotes the right hand side. Green (dot) line is the phase boundary between homogeneous phase and chiral restored phase.Black dot denotes the CEP(LP) and black square denotes the "LP" respectively.}
\label{ddcdw2}
\end{figure}

Minimizing the thermodynamic potential Eq.(\ref{13}) with respect to $\Delta$ and $k$ for each $T$ and $\mu$, we get the QCD phase diagram in Fig.\ref{ddcdw1}. 
Similarly to the case in the chiral limit, the inhomogeneous phase appears between the homogeneous condensate phase and chiral restored phase in the QCD phase diagram. The phase transition is of the first order on the left phase boundary.
On the other hand, since it is a crossover from the inhomogeneous phase to the chiral restored one in the presence of the mass term, there is some ambiguity about the right phase boundary. 
Generally it may be defined as a line where the double-well type function of the effective potential changes to the single-well one 
(the convex condition) \cite{Negele}. 

We simply draw it by invoking the correlation function method \cite{Nakano-Tatsumi} in Fig.~\ref{ddcdw1},\ref{ddcdw2}.
Correlation function method is the exact method which determine the right phase boundary in the chiral limit.
From \cite{Nakano-Tatsumi}, the phase boundary would be fixed by the following conditions;
\begin{align}
1 - 2G \Pi_{\rm{ps}} (q) = 0~~~{\rm{and}}~~~\partial_q \Pi_{\rm{ps}} (q) = 0, \label{corr}
\end{align}
where $\Pi_{\rm{ps}} (q)$ is the correlation function of the pseudo-scalar density with momentum $q$ in the chiral limit.

\begin{figure}[h]
\includegraphics[scale=0.55]{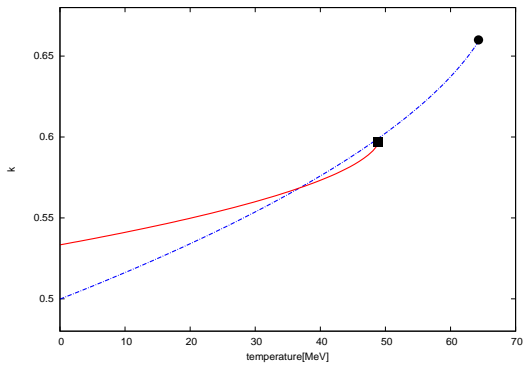}
\caption{Changing of $k$ with increasing T on the phase boundaries of the inhomogeneous phase. Blue (Red) line shows the left (right) phase boundary. Black dot(square) denotes the CEP("LP").}
\label{ddcdw_kt}
\end{figure}

In Fig.\ref{ddcdw2}, we see that the Lifshitz point (LP), which is the triple point of the homogeneous phase, chiral restored phase and inhomogeneous phase, appears near the critical end point (CEP). In fact, numerical calculations show that the wave number becomes smaller monotonously with approaching to LP (see Fig. \ref{ddcdw_kt}  and Eq. (\ref{qfomula}) ).
The critical point given by the correlation function method ("LP") in Fig. \ref{ddcdw2} is located at somewhat different point from CEP, while the generalized Ginzburg-Landau theory (gGL) tells that LP should coincide with CEP \cite{lifshitz}.
Such difference may be attributed to our procedure;
we draw the right phase boundary by using the correlation function method, but it is valid "in the chiral limit".
With switching the current quark mass, the condition Eq.(\ref{corr}) should be properly changed; 
Considering the current quark mass, the Ginzburg-Landau expansion is represented as
\begin{align}
\Omega_{\rm GL} = -\alpha_1 M + \alpha_2 M^2 + \alpha_4 M^4 . \label{GL-NJL},
\end{align}
where the coefficient $\alpha_1$ is the contribution from the current quark mass.
The coefficient $\alpha_2$ represents the inverse of the correlation function in the pseudoscalar channel 
\cite{Nakano-Tatsumi};
\begin{align}
\alpha_2 = \frac{1}{2} \Gamma_{\rm ps}^{-1} (q) = \frac{1}{2} \left( 1-2G\Pi_{\rm ps} (q) \right) .
\end{align}
From Eq. (\ref{GL-NJL}), we can extract the convex condition; $4\alpha_1\alpha_4 = \alpha_2^2$, which determine the right phase boundary after imposing the minimizing condition with $q$.
In the low temperature region, the constituent mass of the restored phase is sufficiently small and we can drop $\mathcal {O} (M^4)$ and set $\alpha_1 \sim 0$, which is just Eq.(\ref{corr}).
Near CEP, on the other hand, where the constituent mass is large and we should consider $\mathcal {O} (M^6)$ in the GL expansion, which ensures the coincidence of LP and CEP. Thus the condition should be changed to describe the CEP region more correctly.
Further discussion would be given elsewhere \cite{full}.

Next we demonstrate the change of the properties of the inhomogeneous phase by changing the value of modulus along the phase boundaries.
In Fig.\ref{ddcdw_kt}, we see that the modulus $k$ monotonously increases or the wave number $q$ decreases 
with increasing temperature toward the critical point. 
This feature can be understood by the equation of motion of $\theta(z)$ Eq.(\ref{9}). 
In our results, the wave number  $q$ decreases monotonously as temperature increases. 
Since the wave number is much larger than the pion mass in the low temperature region, the first term of Eq.(\ref{9}) is dominant to make the inhomogeneous structure not so deformed from the original DCDW.
Near the critical point, however, the wave number becomes even smaller than the pion mass, so that the second term of Eq.(\ref{9}) becomes important and the condensates  have been largely deformed.

\begin{figure}[h]
\includegraphics[scale=0.2]{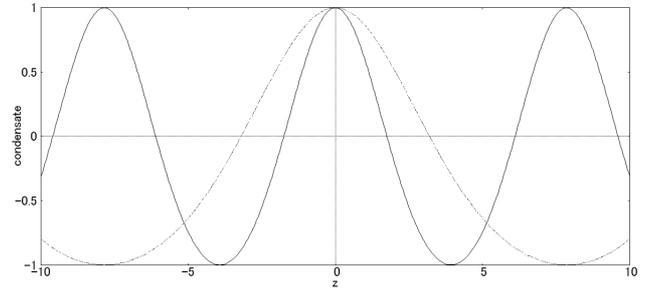}
\caption{Profile of the scalar condensate $\langle{\bar \psi}\psi \rangle$. Solid line shows the value in the low temperature region, while Dash line the value near CEP. The amplitude of $\langle{\bar \psi}\psi \rangle$ is rescaled.}
\label{condensate}
\end{figure}

Finally we see the change of the profile of the chiral condensates.
Fig.\ref{condensate} shows that the scalar condensate oscillates almost periodically in the low temperature region, while it becomes unevenly distributed  near the critical point.
This uneven distribution can be understood as the consequence of the deformation of the chiral circle; the chiral circle is no more flat in the presence of the mass term: the mass effect makes the absolute minimum  on the chiral circle ($\sim -1$ in Fig.\ref{condensate} ), so that the condensate favors to stay around this minimum. 



In this paper, we consider the effect of the current quark mass to the inhomogeneous chiral phase with the DCDW-type structure.
Since there is no exact solution for the DCDW-type structure, we propose a variational method to this problem and figure out some features of the modified DCDW state with the current quark mass.
The equation of motion for the chiral angle is given by the SG equation and the second term in this equation implies the effect of the current quark mass.
Because the current quark mass deforms the chiral circle, the function form of the condensates has been modified from DCDW qualitatively; the chiral angle exhibits  even the kink-like form near the critical point. 
Minimizing $k$ and $\Delta$, we have studied the QCD phase diagram on the $T-\mu$ plane and found that the inhomogeneous phase appears even if the current quark mass is switched on. 
In the low temperature region, modulus $k$ is small so that the deformation from DCDW is small, while $k$ becomes large then the structure 
is much deformed near the critical point.
Finally we discuss the validity of the approximation in our framework. In the well developed phase, where $q$ is rather large or $k$ is small, the spatial modulation of $|d\theta/dz|$ is small and our assumption should be legitimate. 
One may be worried about the large deformation ($k\sim 1$) near the critical point, while our results look to well describe it. Note that the phase with the kink-like solution is equivalent with the uniform phase \cite{BDT}; actually we can immediately see that $q\rightarrow 0$ and $\Omega_{\rm SB}\rightarrow 0$ in Eq.~(20). This is why our ansatz can describe the critical point correctly.   

Thus we have seen that our approach works well in the vicinity of the critical point or the phase boundary as well besides the fully-developed DCDW phase.
However, the numerical calculation may be possible to find the solution corresponding to the deformed DCDW, by solving the Hartree equations.
It should be then interesting to check the validity of our approach by comparing our result with them.
This subject is left as a future work.

\bibliographystyle{model1b-num-names}
\bibliography{<your-bib-database>}







\end{document}